# Drawing a parallel between the trend of confirmed COVID-19 deaths in the winters of 2022/2023 and 2023/2024 in Italy, with a prediction

**Roccetti Marco***

Department of Computer Science and Engineering, University of Bologna, Bologna, Italy

* **Correspondence:** Email: marco.roccetti@unibo.it; Tel: +393920271318.

**Abstract:** We studied the weekly number and the growth/decline rates of COVID-19 deaths of the period October 31, 2022 – February 9, 2023, in Italy, finding that that COVID-19 winter wave reached its peak during the three holiday weeks from December 16, 2002 to January 5, 2023, and it was definitely trending downward, returning to the same number of deaths of the end of October 2022, in the first week February 2023. During this period of 15 weeks, that wave caused a number of deaths as large as 8,526. Its average growth rate was +7.89% deaths per week (10 weeks), while the average weekly decline rate was -12.32% (5 weeks). At the time of writing of this paper, Italy was experiencing a new COVID-19 wave, with the latest 7 weekly bulletins available at that date (October 26, 2023 – December 14, 2023) showing that the deaths had climbed from 148 to 322. The weekly growth rate had risen by +14.08% deaths, on average. In the hypothesis that this 2023/2024 wave will have a total duration similar to that of 2022/2023, with a comparable extension of both the growing period and of the decline period and similar growth/decline rates, a prediction was cast regarding the number of COVID-19 deaths of the period end of October 2023 – beginning of February 2024 which should be near 4100. A preliminary assessment of this forecast, based on 11 of the 15 weeks of the period, has already confirmed the accuracy of the approach.

## 1. Introduction

During the 2022/2023 autumn-winter season in Italy, we have observed a typical recurring pattern in the curve of the daily confirmed COVID-19 deaths. In fact, the number of daily deaths began to increase after the schools' opening (end of September 2022) [1]. Then this increasing trend stabilized around a quite high value of 200-300 weekly deaths by the beginning of October 2022. Afterwards, winter came in Italy with a combination of meteorological, social and environmental factors, including: i) more rigid temperatures, ii) several holiday periods, among which All Saints'/Dead festivities (November, 1-2), the Celebration of the Immaculate Conception (December, 8) the Christmas and New Year festivities (December, 24 – January, 6), and iii) people gatherings in closed spaces.

It is well known that within a gathering, to which many people take part, the presence of infected persons is highly likely. This can ignite infections inside the gathering, as the spread of COVID-19 occurs via airborne particles and droplets. In addition, in a gathering, infected individuals have more opportunities to come at close quarters with other people, with an infection rate resulting higher than in other situations. Not only that, but after a gathering event, the participants return to their lives, thus contributing to the spread of the epidemic throughout society [2,3].

As a results of this series of festivities and consequent people gatherings, the curve of daily confirmed COVID-19 deaths started a new climb, somewhere around the end of October, 2022, up to a peak that was registered around the beginning of January 2023, after exactly ten weeks since the climb has begun, and with a weekly number of deaths as large as 800 registered in those three holidays weeks. After that peak, a quite rapid downward trend began which stabilized in almost five weeks, returning approximately to the same number of COVID-19 confirmed deaths that were experienced when that wave had started (end of October: 200-300 deaths).

In the end, during this observed period of fifteen weeks, that 2022/2023 COVID-19 wave caused a number of deaths as large as 8,526, with: a) an average number of weekly deaths equal to 568, b) an average growth rate of +7.89% deaths per week (for ten consecutive weeks), and c) an average decline rate of -12.32% (for five consecutive weeks). All those data are officially available at the web site of the Italian Ministry of Health that publishes, on a regular weekly basis, its epidemiological bulletins [4].

The situation we have summarized has a visual counterpart in Figure 1, where the daily new confirmed COVID-19 deaths (per million people) in Italy is plotted over the period that goes from the end of August 2022 to the mid of April 2023. In particular, the two leftmost black thin vertical bars in Figure 1 demarcate the period when the schools opened (mid of September – beginning of October, 2022). The green thick bar marks the period when All Saints'/Dead festivities occurred (beginning of November 2022). The yellow thick bar, instead, marks the period corresponding to the Celebration of the Immaculate Conception (December, 8) and finally the blue thick bar delineates the beginning of the period of the Christmas and New Year festivities. The curve also shows, very clearly, all the increasing and decreasing trends we have described before. The curve represented in Figure 1, with all its data, has been drawn using the open data source made available by the Our World in Data initiative [5].

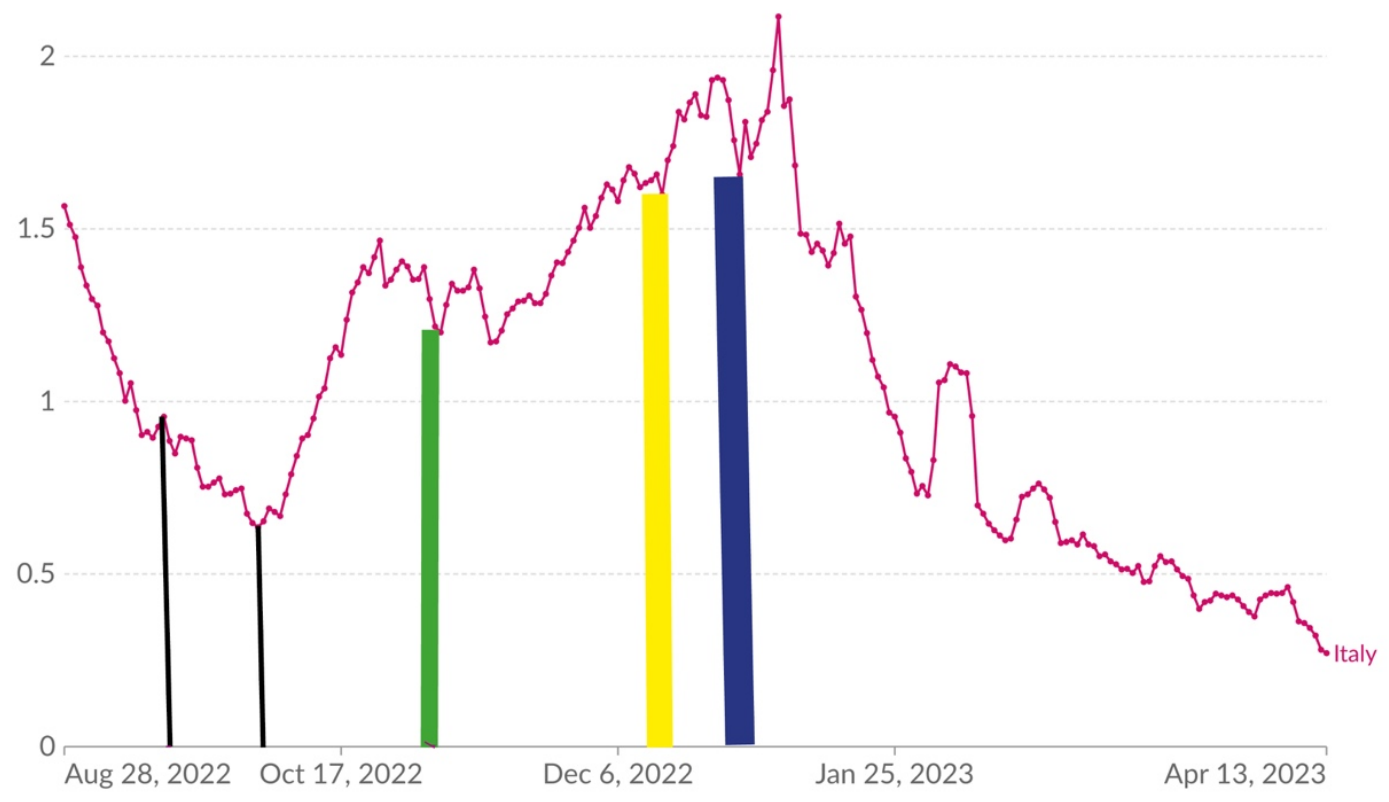

**Figure 1.** Daily new confirmed COVID-19 deaths per million people in Italy. On the y-axis the 7-day rolling average of weekly deaths relative to Italian population. On the x-axis, the period: end of August

2022 – mid of April, 2023. Sources: Our World in Data (2023).

The aim of this paper has bee to analyze the weekly number and the growth/decline rates of the COVID-19 deaths of the 2022/2023 COVID-19 winter season, along with the periods in time when it peaked and then declined down thus reaching a stable situation. All this effort has been done with the intent to project those numeric figures onto the Italian COVID-19 situation which had experienced a resurgence in cases and deaths in the winter season 2023-2024, starting again at end of October 2023.

At the time of writing of this paper (December 15, 2023), in fact, we knew that in the period from October 26, 2023 to December 14, 2023, comprised of seven consecutive weeks, the COVID-19 deaths have climbed from 148 to 322 on a per week basis, with an average weekly growth rate of +14.08%, as confirmed by the corresponding seven Ministry bulletins [4].

The final objective of this study was to cast a prediction of the death toll Italy will have paid to the 2023-2024 winter COVID-19 wave (from the last week of October 2023 to first week of February 2024), under the hypothesis that it would have had a duration similar to that of 2022/2023 (fifteen weeks) with comparable values of: i) the duration of the growing period (ten weeks) and of the decline period (five weeks) and ii) the growth/decline rates.

In simple words, we have tried to use a method based on the concept of the so-called *predictive analogy* where the fundamental idea is that future events and outcomes can be forecast by comparing them to past experiences and patterns. This approach requires drawing parallels between similar situations to predict an event or a trend [6, 7]. We know very well that this approach is subject to many limitations and runs the risk of failure. Nonetheless, we have decided to use neither the traditional COVID-19 spread indicators of the classical SIR/SEIR models, nor alternative but simpler regression models [8], because recent studies have shown that more basic and intuitive metrics can establish a better correlation, than traditional indicators, with both the number of hospitalized patients and deaths during a COVID-19 epidemic [9, 10, 11, 12, 13, 14]. Even if epidemiologists may remain skeptical about the use of those basic metrics for carrying out a real time surveillance, we are confident that our approach can still be useful to predict deaths over an adequate period of time, essentially because it is not influenced by the large number of asymptomatic carriers which remain still unknown and have often contributed in the past to the failure of traditional epidemiological methods used to modelling the COVID-19 diffusion.

In conclusion, we have used analogy, in this context, to compare two analogs (the two COVID-19 winter waves occurred in two consecutive years in Italy), and then to cast a prediction on the number of deaths, based on the observed similarities and parallels. The numbers of this prediction have been anticipated in the Abstract and will be explained in detail in Section of Results, along with a preliminary assessment that seems to confirm the validity of the approach.

The remainder of this paper is the following. In the Section termed Materials and Methods we illustrate where our data come from, along with the methodology that we have used to analyze them in order to cast our prediction. In a subsequent Section, termed Results, we illustrate the Results we have obtained and then in a Section, termed Discussion, we illustrate both the advantages and the limitations of our approach. Finally, a Section termed Conclusions terminates our paper.

## 2. Materials and methods

### *2.1. Sources of data*

In this Section, we provide sufficient details to allow readers to replicate our results. To begin, it is worth revealing the source of data we have used for our study. We have already anticipated that they have been all drawn mostly by one official source: The Italian Ministry of Health. More specifically, this Institute issues, on a per week basis, a comprehensive bulletin where the number of confirmed COVID-19 deaths are published [4]. As an alternative source of data, in the case when the primary web site of the Ministry of Health is unavailable, it is also possible to refer to the official web site of the Italian Broadcast Television which simply provide a mirror of those official data [15]. Obviously, even if we know that, because of varying protocols and decisions in the attribution of the cause of death, the number of confirmed deaths may slightly vary over time, these bulletins provide the official number of the COVID-19 deaths registered by the Italian health authorities.

In summary, the public repositories with all data used in this paper are respectively available at:

https://www.salute.gov.it/portale/nuovocoronavirus/archivioBollettiniNuovoCoronavirus.jsp (Italian Health Ministry), and

https://www.rainews.it/ran24/speciali/2020/covid19/ (Italian Broadcast Television).

At the time of writing of the first draft of this paper (December 15, 2023), the main source of data represented by the Italian Ministry of Health was comprised of 59 weekly bulletins, starting from the week: October 28 - November 3, 2022 until the latest week: December 6 – December 14, 2023. In all these bulletins, a Table was reported, filled with a weekly update of a variety of COVID-19 indicators. Beyond many others, the number of the weekly confirmed deaths is clearly shown.

### *2.2. Method of analysis*

Said about the source of data, the method of analysis we have used has been the following. Inspired by the information portrayed in Figure 1 above, where the beginning, the end, and the peak of the COVID-19 winter wave of season 2022/2023 were reported, we decided to use a method based on "*historical analogies*" to analyze the available data, as well as to cast a new prediction. A few words are in order to introduce this method.

As already anticipated in the Introduction, this kind of model has roots in several scientific fields, from product marketing to software adoption [6, 7], with the unique idea of drawing parallels between phenomena and then modeling one on the imitation of the other [16]. The decision if two phenomena follow the same behavior depends on their similarities/differences and can be taken by an expert, or by a pool of experts. In this sense, it is very similar to the Delphi method, which is a systematic process of forecasting, using the collective opinion of recognized experts [17, 18]. It is not by chance that the technique of developing consensus among experts using the Delphi approach has gained acceptance in diverse fields of medicine. Obviously, the obtained results do not allow the use of traditional method of hypothesis testing (e.g., p-values) as those values do not come from a distribution to test but should be intended as point of references, as a trend that the new phenomenon is supposed to follow.

A recent case of use of this methodology for forecasting is in the field of technology, where active attempts have been developed to predict life cycles of given technology products, based on the observation of what was written in news, papers, patents and other publications, related to that kind of technology, and published some time before. Applied it to a case for estimating the Korean Plug-in Hybrid Electric Vehicle market, historical analogy has emerged as a promising new dimension for forecasting [19].

Along this line of sense, many researchers are (re)considering analogies as a good method for empirically predicting outcomes, under specific circumstances. We will discuss about the circumstances that make this method profitable in our case in the Discussion Section.

*2.3. Implementation of the method*

After the above abstract description of the "*forecast by analogy*" methodology, we come to the practical procedure we have implemented, which is as follows: first, we downloaded from the web site of the Italian Ministry all the fifteen bulletins corresponding to the fifteen weeks of interest for the winter season of year 2022/2023. They were the following: 1) October 28 – November 3, 2022; 2) November 4 – November 10, 2022; 3) November 11 – November 17, 2022; 4) November 18 – November 24, 2022; 5) November 24 – December 1, 2022; 6) December 2 – December 8, 2022; 7) December 9 – December 15, 2022; 8) December 16 – December 22, 2022; 9) December 23 – December 29, 2022; 10) December 30, 2022 – January 5, 2023; 11) January 6 – January 12, 2023; 12) January 13 – January 19, 2023; 13) January 20 – January 26, 2023; 14) January 27 – February 2, 2023; 15) February 3 – February 9, 2023.

From those fifteen bulletins, we extracted the fifteen values of the weekly deaths and then we computed the percentage of the change for each consecutive pair of values of the growth/decline rates of the series, comparing the most recent value in the series with the previous one. In particular, we called these values: P*ercentage* c*hange of the* w*eekly growth factor* when we were examining the ascending phase of the COVID-19 wave, while we called that value P*ercentage* c*hange of the* w*eekly decline factor* when the descending phase of that wave was under examination. More specifically, both the aforementioned values were calculated according to the following formula.

$$Percentage\ change\ of\ the\ weekly\ growth \backslash decline\ factor = ((\left(\frac{ND(i)}{ND(i-1)}\right) * 100) - 100)\% \quad (1)$$

The meaning of the variables *ND(i)* and *ND(i-1)* is as follows. *ND(i)* indicates the number of deaths of a given week *i*, while *ND(i-1)* represents the number of deaths registered during the week the precedes *i*. Obviously, the sign of the result will be naturally assigned as either positive or negative depending, respectively, if *ND(i) is* either greater than *ND(i-1)* or not.

All this has produced an ordered series of fourteen different values of the percentage changes of the *growth/decline factors,* corresponding to the fifteen weeks under study. Of the first nine of these values, we computed the average, obtaining the average percentage of the weekly growth rate of the deaths during the ten consecutive weeks when the curve of the year 2022/2023 was in its increasing phase. Similarly, we used the last five values returned by the Formula above, to get the average percentage of the weekly decline factor of the deaths occurred during the five consecutive weeks when the curve of deaths of 2022/2023 had declined down to its initial baseline (value of the baseline: almost 200-300 weekly deaths).

At that point, we repeated the same procedure above for the seven bulletins issued by then Italian Health Ministry, relative to the seven weeks of the season 2023/2024, available at the time of writing of the first draft of this paper (December 15, 2023). Specifically, we started by downloading the bulletins of the following weeks: 1) October 26 – November 1, 2023; 2) November 2 – November 8, 2023; 3) November 9 – November 15, 2023; 4) November 16 – November 22, 2023; 5) November 23 – November 29, 2023; 6) November 30 – December 6, 2023; 7) December 7 – December 14, 2023.

From these seven bulletins, we extracted the seven actual values of the weekly COVID-19 deaths and then we computed the percentage of the changes of the growth factor, based on the same Formula 1 above. We obtained a series of six values of the weekly percentage changes of the growth rate of the deaths, relative to the seven consecutive weeks of the 2023/2024 infection (which was still in its increasing phase, at the date of December 15, 2023). Finally, we computed the average of these six values.

We are near to cast our prediction. To summarize, we remind that we could count on: the series of the percentage values of the changes of the growth/decline factors for the year 2022/2023, comprised of fourteen vales, plus: 1) an average percentage of the first nine values of the series above (we call this value: *average weekly growth trend*), relative to the first ten consecutive weeks when the curve of the deaths of year 2022/2023 was climbing, and 2) an average percentage of the last five values of the series above (called *average weekly decline trend*) relative to the last five weeks when the curve of deaths of year 2022/2023 was definitely in its downward phase. Finally, we also had all the fifteen actual values of the number of deaths experienced during the entire period of fifteen weeks mentioned above.

Of year 2023/2024, instead, at the time of writing the first draft of our paper (15 December 2023), we had only seven actual values of the number of deaths experienced during the period of seven weeks, from the end of October 2023 till mid December 2024. Moreover, we had a series of only six percentage values of the changes of the growth factor (which are relative to the seven consecutive weeks of the current season 2023/2024, till mid December 2024). Obviously, we also have the average of the values of this series (called *average weekly growth trend*).

The problem to solve now is how to predict the eight remaining actual weekly values of the COVID-19 deaths, for the period which goes from the mid of December 2024 to the beginning of February 2024, thus yielding a full series of fifteen values.

There are two alternatives to achieve this result, based on the concept of analogy and built on two different sets of assumptions. We will term the first set of assumptions as Assumption A and the second one as Assumption B.

In fact, under the conditions of Assumption A, we take as valid the following hypotheses: i) the COVID-19 wave of season 2023/2024 will have a duration similar to that of 2022/2023 (fifteen weeks), ii) a comparable extension of both the growing period (ten weeks) and of the decline period (five weeks) will occur, and iii) we will use the *average weekly growth trend* of year 2022/2023 and the *average weekly decline trend* of year 2022/2023 in order to calculate, respectively, 1) the predicted values of the COVID-19 deaths for the three remaining weeks of season 2023/2024, still supposed to be in its ascending phase, and 2) the predicted values of the COVID-19 deaths for the last five weeks of the descending phase of season 2023/2024.

Under the hypothesis of Assumption B, conditions i) and ii) still hold unchanged, but with a difference in condition iii) consisting in the idea to use an estimate of the *average weekly growth trend* of year 2023/2024, computed with only the first seven weeks of season 2023/2024 (from end of October to mid of December, 2023), as a basis for prediction of all the remaining eight values of that season (mid of December 2023 – beginning of January 2024).

With this change in condition iii), the additional problem emerges of how estimating the *average*

*weekly decline trend* for the last five weeks of the descending phase of the COVID-19 wave of season 2023/2024, only starting from the estimation of the *average weekly growth trend* of season 2023/2024, (computed, in turn, on the basis of the values relative to the first seven weeks till mid December 2023). To solve this problem, one can adopt the following strategy.

We take: i) the (percentage) *average weekly growth trend* of year 2022/2023 (call it *X*%, for the sake of brevity), and ii) the (percentage) *average weekly decline rate* of year 2022/2023 (call it *Y*%, for the sake of brevity), and then we calculate the percentage change between *X* and *Y*, using the Formula 2 below:

$$Percentage\ change\ between\ X\ and\ Y = \Big(\big((abs(Y/X)) * 100\big) - 100\Big)\% \qquad (2)$$

At that point, we know that on average in the year 2022/2023 the change between the growth (*X*%) and decline (*Y*%) trends has been equal to a given percentage quantity computed with the above Formula 2, and equal to, say, *W*%, where *W* will be naturally either positive or negative depending respectively if *abs(Y)* is greater than *X* or not.

The idea is to exploit this value *W*% and the estimate of the (percentage) *average weekly growth trend* of year 2023/2024 we have already computed with only the first seven weeks of year 2023/2024 (call it *Z*%). To obtain an estimate of the *average weekly decline trend* of year 2023/2024, we can proceed as follows.

With *W* and *Z*, we can compute a new value *K*% = *((Z * W) / 100)%.* Roughly speaking, *K* represents the amount with which we should increment *Z* to obtain a decline trend analogous to that experienced in the previous season. Consequently, with *K*% we can obtain an estimate of the (percentage) *average weekly decline trend* for year 2023/2024 (call it *Q*%), by applying the following formula:

$$Q\% = (Z + K)\% \qquad (3)$$

Now, we can use the estimate *Z*% of the (percentage) *average weekly growth trend* of season 2023/2024 to all the three weeks from December 22, 2023 to January 4, 2024 and obtain the three predicted values of the COVID-19 deaths for those three weeks.

Then, we can apply the *(- Q%)* of the (percentage) *average weekly decline trend* of season 2023/2024 to the last five weeks of the series, corresponding to the period from January 5 to February 8, 2024, thus obtaining the predicted values of the COVID-19 deaths for those five weeks that complete the period of interest. As intended since the beginning of this study, these eight predictions cover the entire period from December 15, 2023 to the beginning of February 2024.

All the results that can be obtained with the Methods we have discussed are presented in the next Section

#### 2.3.1. Ethics approval of research

This study uses publicly available, aggregated data that contains no private information. Therefore, ethical approval is not required.

## 3. Results

We now present the results we have obtained using the data and the methods described in the previous Sections. They are all summarized in Table 1.

The first fact to notice is that in the first column of Table 1 all the fifteen weeks of interest are listed, both for years 2022/2023 and 2023/2024. The second column reports the number of confirmed weekly deaths due to COVID-19, again for seasons 2022/2023 and 2023/2024. The values we have predicted with our model are shown with the Bold/Italic font in the Table. Specifically, on the left of the slash symbol (/) the predictions are reported which have been made under the conditions of Assumption A, while on the right the predictions can be found achieved under the conditions of Assumption B.

In the third column, the (percentage) changes of the weekly growth/decline factors/trends are shown. Again, we show the predicted values in Bold/Italic. Also in this case, the predictions of Assumption A and of Assumption B are respectively shown on the left and on the right of the slash symbol (/). As to the method that builds on Assumption B, it is worthwhile mentioning that it has returned the following values: $Z\% = 14.08\%$ and $- Q\% = - 22.1\%$.

Obviously, the most important result of this study is the cumulative number of predicted deaths computed on the basis of the methods described in the previous Section. It amounts, respectively, to: 4,166 and 4,123 COVID-19 deaths (as shown in the last row of Table 1), depending on the use of the conditions of, respectively, Assumption A and Assumption B. A detailed discussion on the relevance of this result follows in the Section Discussion.

**Table 1.** Number of COVID-19 confirmed weekly deaths and changes of the growth/decline trends. In Bold and Italic, the predicted values. Predicted values based on Assumption A are on the left of the slash (/). Predicted values based on Assumption B are on the right of the slash symbol. Predictions start from the 8th week.

| **# Weeks**<br><br>**First line: 2022/2023**<br><br>**Second line: 2023/2024** | **Weekly deaths:** (actual/**predicted)**<br><br>**First line: 2022/2023**<br><br>**Second line: 2023/2024** | **Changes:** (actual/**predicted)**<br><br>**First line: 2022/2023**<br><br>**Second line: 2023/2024** |
|---|---|---|
| 1: Oct 28/Nov 3, 2022<br>1: Oct 26/Nov 1, 2023 | 1: 411<br>1:148 | null |
| 2: Nov 4/Nov 10, 2022<br>2: Nov 2/Nov 8, 2023 | 2: 549<br>2:163 | 2: +33.58%<br>2: 10.13% |
| 3: Nov 11/Nov 17, 2022<br>3: Nov 9/Nov 15, 2023 | 3: 533<br>3:192 | 3: -2.91%<br>3: +17.79% |
| 4: Nov 18/Nov 24, 2022<br>4: Nov 16/Nov 23, 2023 | 4: 580<br>4: 235 | 4: +8.81%<br>4: +22.39% |
| 5: Nov 24/Dec 1, 2022<br>5: Nov 23/ Nov 29, 2023 | 5: 635<br>5: 291 | 5: +9.48%<br>5: +23.83% |
| 6: Dec 2/Dec 8, 2022<br>6: Nov 30/Dec 6, 2023 | 6: 686<br>6: 307 | 6: +8.03%<br>6: +5.50% |
| 7: Dec 9/Dec 15, 2022<br>7: Dec 7/Dec 14, 2023 | 7: 719<br>7: 322 | 7: +4.81%<br>7: +4.88% |
| 8: Dec 16/Dec 22, 2022<br>***8: Dec 15/Dec 21, 2023*** | 8: 798<br>***8: 347/367*** | 8: +10.99%<br>***8: +7.89/14.08%*** |
| 9: Dec 23/Dec 29, 2022<br>***9: Dec 22/Dec 28, 2023*** | 9: 706<br>***9: 374/419*** | 9: 11.53%<br>***9: +7.89/14.08%*** |
| 10: Dec 30, 22/Jan 5, 23<br>***10: Dec 29, 23/Jan 4, 24*** | 10: 775<br>***10: 404/478*** | 10: +9.77%<br>***10: +7.89/14.08%*** |
| 11: Jan 6/Jan 12, 23<br>***11: Jan 5/Jan 11, 24*** | 11: 576<br>***11: 354/372*** | 11: -25.68%<br>***11: -12.32/-22.01%*** |
| 12: Jan 13/ Jan19, 23<br>***12: Jan 12/Jan 18, 24*** | 12: 495<br>***12: 310/290*** | 12: -14.06%<br>***12: -12.32/-22.01%*** |
| 13: Jan 20/Jan 26, 23<br>***13: Jan 19/ Jan 25, 24*** | 13: 345<br>***13: 272/226*** | 13: -12.66%<br>***13: -12.32/-22.01%*** |
| 14: Jan 27/Feb 2,23<br>***14: Jan 26/Feb 1, 24*** | 14: 439<br>***14: 238/176*** | 14: +27.25%<br>***14: -12.32/-22.01%*** |
| 15: Feb 3/Feb 9, 23<br>***15: Feb 2/Feb 8, 24*** | 15: 279<br>***15: 209/137*** | 15: -36.45%<br>***15: -12.32/-22.01%*** |

| | |
|---|---|
| End of period of investigation | 15-week period |
| 2022/2023: Cumulative number of deaths | 8,526 |
| ***2023/2024:Cumulative number of predicted deaths*** | ***4,166-4,123*** |

*3.1. A preliminary assessment*

The design of our model, the corresponding predictions and the writing of the first draft of this manuscript date back to December 15, 2023, when we were in the seventh week (out of fifteen) of our period of investigation. Today, date of submission of this paper (January 19, 2024), it has been four weeks since that moment. This has allowed us to compare some of the predictions we made with the actual values, communicated by the Italian Health Ministry [4, 15], of most recent four weeks (namely, weeks: 8, 9, 10, 11 of Table 1).

We have conducted a comparison between our predictions and the actual values, in terms of the cumulative number of deaths occurred, considering two different situations. The former where the comparison is limited to just those most recent four weeks (weeks: 8, 9, 10, 11 of Table 1) and the latter where the entire period is taken into account since its beginning (last week of October 2023 to our current days), namely, weeks: 1, 2, 3, 4, 5, 6, 7, 8, 9, 10, 11 of Table 1.

In the following Table 2, the results of this comparison are shown, with respectively: the cumulative number of actual deaths in the second column of Table 2, the cumulative number of predicted deaths (under Assumption A) in the third column, and finally the cumulative number of predicted deaths (under Assumption B) in the fourth column of Table 2. Along with the predicted number of deaths, we have computed also the percent error (PE) of each prediction.

Observing Table 2, we can notice that the predictions cast under Assumption A are currently very accurate, with a PE in the range 1.6-3.4%, for both of the situations analyzed (i.e., 4 vs 11 weeks). Less accurate, but still acceptable is the prediction made under the conditions of Assumptions B, whose PE is in the range 6.7-14.4%. Generally, in favor of both these predictions (i.e., Assumption A and B) is the fact that this preliminary assessment has been conducted with the four weeks considered the most crucial for the impact of COVID-19 on the population, as they coincide with the period of the Christmas and New year festivities. Moreover, if we consider that: i) we have had so far 3088 COVID-19 deaths, and ii) we are four weeks away (remaining weeks: 12, 13, 14, 15) from the end of the 15-week period of interest, then the prediction of having a total amount of COVID-19 deaths near 4100 is a very realistic guess.

**Table 2.** Cumulative number of actual COVID-19 deaths in Italy (second column), Cumulative number of predicted COVID-19 deaths under Assumption A (third column), Cumulative number of predicted COVID-19 deaths under Assumption B (third column).

| | | | |
|---|---|---|---|
| **4 weeks period (Dec 14, 2023 – Jan 11, 2024): weeks 8-11** | Cumulative number of actual deaths: **1430** | Cumulative number of predicted deaths (Ass. A): **1479** | Cumulative number of predicted deaths (Ass. B): **1636** |
| **11 weeks period (Oct 26, 2023 – Jan 11, 2024): weeks 1-11** | | **PE: 3.4%** | **PE: 14.4%** |
| | Cumulative number of actual deaths: **3088** | Cumulative number of predicted deaths: **3137** | Cumulative number of predicted deaths: **3294** |
| | | **PE: 1.6%** | **PE: 6.7%** |

## 4. Discussion

Analogy is often used for comparing two analogs, based on obvious similarities between the elements that comprise the analyzed phenomena. As such, it assumes that the two phenomena of interest follow almost the same behavior, on the basis of which similar outcomes can be predicted. Obviously, the degree of similarity, on which analogy is based, is greatly influenced by the considered attributes. Needless to say, these are exactly both the strengths and the limitations of our approach. In the specific case of our study, in fact, similarities can be found, mainly, in: i) the winter period of investigation, with its meteorological conditions, ii) the behavior of people during the winter holidays, with their typical attitude toward gatherings, iii) the total number of administered vaccinations which is not greatly changed in the two periods of interest (as an example, consider that the number of boosters administered during the period October-December 2023 were as few as 1,500,000 [20], while the total number of fully vaccinated people were almost 48,000,000 in Italy since October 2022 [21]), iv) although new variants are an expected part of the evolution of any virus, Omicron and its sub-variants had ranked as the predominant SARS-CoV-2 strains for both the winter seasons of interest, making the situation enough homogenous in both those two winter seasons [22].

Another reason in favor of use of the methodology adopted in our study is that it is based on the repetition of similar events in recent history. In our case, seasonal holidays and people gatherings seem to be a perfect instance of this concept of repetition over different years. It is not by chance, in fact, that also other authors had already noticed this fact maintaining that in Italy the socio-medical structure developed to contain the pandemic "*broke down in the summer as Italians went on holiday….*" [23].

Obviously, a strong limitation in using historical analogies to predict outcomes is concerned with the time scale used for the investigation. In fact, if the time scale is too short, the available amount of information (in our case, the number of weeks with relative deaths) could be insufficient to draw a parallel between different seasons. On the other hand, if the time scale becomes too long, this would mean that too many events, confounding factors and unexpected accidents could play a role, thus making the model based on analogies too complex to be managed. Coming to our case, the three-months period we have investigated seems to be adequate, since it is characterized by the repetition of the already mentioned seasonal events over two different years. Instead, extending over a longer term would make complex a careful application of the analogies.

All this said, from the perspective of the parallel we have drawn, we are persuaded that our analogical study can be a considered as a useful source of knowledge, especially if the values we have predicted are thought of as simply indicating a trend or some kind of reference values. This means that we expect that our analysis may err in predicting the precise values, with our predicted values which could be higher or lower than the actual ones, with equal probability [24]. Nonetheless, if we try to find sense to our predictions, we are sure that even thinking and discussing analogies may generate useful (and sometimes crucial) information.

Not only that but it is worth noticing that the preliminary assessment we have conducted, based on a comparison between the cumulative number of actual COVID-19 deaths and the cumulative number of deaths predicted on the basis of our methodology, has already shown that our prediction is by now accurate, with a high probability of a percent error contained approximately below the value of 7%.

We also think that our findings and their implications should be discussed in a broadest context. For example, the total number of COVID-19 deaths we have predicted for the period November 2023 – January 2024 should be interpreted as a general sign of concern, with health authorities that should urge the public to adhere always to reasonable health guidelines and to get tested for COVID-19 if symptoms arise.

Of final interest are also two other comparisons. The first one regards our prediction and the actual number of confirmed deaths in the same time period of year 2022/2023. There is a sensible difference between the two values: 8,526 vs 4,100, thus confirming the downward trend of the epidemic in Italy. Nonetheless, it is useful to remember, as an exemplar case from the past, that a great concern was expressed by the Italian health authorities, in the winter period of year 1969-1970, when the so-called Asian flu (a subtype of the H3N2 flu virus) caused as many as 5,000 deaths in Italy [25, 26]. This argument is in favor of the fact that if our predicted number of deaths will actually occur, Italians need to adopt a cautious attitude, using common sense and persevering with good practices and vigilance, while avoiding to give in to an excessive feeling of paranoia.

## 5. Conclusions

We have drawn a parallel between the two winter periods of years 2022/2023 and 2023/2024, worried by the COVID-19 infection trend that has rose up again in Italy, starting at the end of October 2023. Based on the concept of *forecast by analogy*, we have tried to estimate the possible number of deaths that could have happened in Italy in the period end of October 2023 – beginning of February 2024. We found that this number could be near 4,100. It is worth noticing that the method we used, based on *predictive analogies*, has proven to be both adequate and innovative. It has been adequate because the preliminary assessment we have conducted on 11 of the 15 weeks has already anticipated the high accuracy of the prediction, showing a percent error in the range of approx. 1.5-7%, (regarding the cumulative number of deaths). It is highly likely that this level of accuracy (with a percent error not much different from 7%) will be confirmed in the end, given that a very few weeks are still missing, and considering that the observed trend, as correctly predicted, is already in its downward phase. It is also innovative because in this specific field, to the best of our knowledge, it is the first time it is used. Nonetheless, it is important to remind that it should be be applied only if one is sure that the similarities are much more than the differences when the two phenomena are compared.

**Author Contributions**

Not applicable: single author.

**Funding**

This research received no external funding

**Informed Consent Statement**

Not applicable: neither humans nor animals nor personal data are being involved in this study.

**Conflict of interest**

The author declares there is no conflict of interest.